\documentclass[aps,prl,twocolumn,groupedaddress]{revtex4}
\usepackage{graphicx}
\def\msun {{\rm M}_{\odot}}
\def\mpc {h^{-1}~{\rm{Mpc}}}

\newcommand{\bm}[1]{\mbox{\boldmath$#1$}}

\begin{document}

\title{Lensing Magnification and QSO-Galaxy Cross-Correlations: \\ Observations, Theory and Simulations}
\author{Antonio C. C. Guimar\~aes}
\email[]{aguimaraes@astro.iag.usp.br}
\altaffiliation{Affiliated to Durham University at the date of the talk presented at the meeting ``100 years of relativity -- international conference on classical and quantum aspects of gravity and cosmology'', 
held in S\~ao Paulo, Brazil, from August 22 to 24, 2005.}
\affiliation{Departamento de Astronomia, IAG,
Universidade de S\~ao Paulo}
\date{\today}

\begin{abstract}
We review observations and gravitational lensing theory related to the magnification of background QSOs by intervening overdensities, and the induced cross-correlation between sources and foreground galaxies.
We pay special attention to simulations, and present some preliminary results from high resolution cluster simulations, which show the role of the halo core and substructure on non-linear magnification.
For massive clusters, deviations from the weak gravitational lensing regime are significant on arcmin scales and bellow.
The accumulated knowledge in the field already shows that gravitational lensing magnification is an important astrophysical and cosmological tool.
\end{abstract}
\maketitle

\section{Observations}

If we take two populations of astronomical objects separated by a sufficiently 
large distance (for example galaxies at low redshift and QSOs at large 
redshift) we expect that there will be no physical connection between them, 
and therefore the positions in the sky of the members of these two populations 
would be uncorrelated. 
However this is not the case in reality.

\begin{figure}
  \centering
  \vspace{0.2cm}
  \includegraphics[width=9 cm]{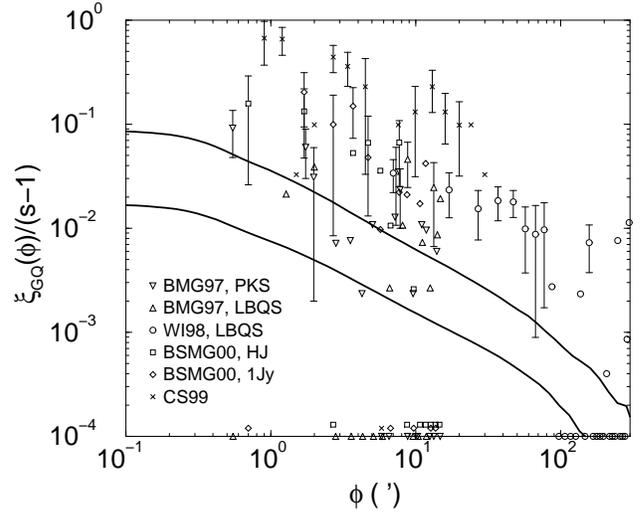}
  \caption{\label{observations} 
    Compilation of some observational determinations of QSO-galaxy cross-correlation. The solid lines show theoretical predictions from analytical calculations assuming weak gravitational lensing, $\Lambda$CDM ($\Omega_m=0.3$,
$\Omega_{\Lambda}=0.7$, $\Omega_b=0.019/h^2$, $n=1$, $h=0.7$,
$\sigma_8=1.0$), and foreground lens populations of power spectrum equal to the APM Galaxy survey (lower curve) and Abell-ACO Galaxy Cluster Survey (upper curve).}
\end{figure}

In fact, several groups have been measuring the cross-correlation between the 
angular positions of objects at high redshits with objects at low redshifts. 
Bellow follows an incomplete, but representative, list. 
Some of the data sets are shown in Figures \ref{observations} and \ref{cross-correl_obs}  ($s$ refers to the double log slope of the cumulative number count as a function of flux for background luminous sources).

-1979, Seldner \& Peebles \cite{seldner-peebles79} found an {\it excess} of galaxies within 15 arcmin of 382 QSOs.

-1988, Boyle, Fong \& Shanks \cite{boyle88,croom99} (CS99 in the legend of Figure \ref{observations}) found an {\it anticorrelation} between faint high-redshift QSOs ($s=0.78$) and low-redshift galaxies from machine measurements of photographic plates.

-1997, Ben\'{\i}tez \& Mart\'{\i}nez-Gonzales \cite{benitez97} (BMG97 in Figure \ref{observations}) found a {\it positive cross-correlation} between 144 radio-loud PKS QSOs ($s=3.5$) and COSMOS/UKST galaxies, and {\it no correlation} when using 167 optically selected LBQS QSOs ($s=2.5$). 

-1988. Williams \& Irwin \cite{williams-irwin98} (WI98 in Figure \ref{observations}) found a {\it strong cross-correlation} between optically selected QSOs ($s=2.75$) from the LBQS Catalog and APM galaxies.

-2001, Ben\'{\i}tez, Sanz  \& Mart\'{\i}nez-Gonzales \cite {BSM01} (BSM00 in Figure \ref{observations}) found {\it positive cross-correlations} between radio-loud quasars from 1-Jy ($s=1.93$) and Half-Jansky ($s=1.42$) samples and COSMOS/UKST galaxies.

-2003, Gaztanaga \cite{gaztanaga03} found a {\it strong positive cross-correlation} between QSOs ($s=1.88$) and galaxies from the SDSS early data release.

-2003 and 2005, Myers et al. \cite{myers03,myers05} found {\it strong negative cross-correlations} (anticorrelations) between $\sim 22,000$ faint 2dF QSOs ($s=0.725$) and $\sim 300,000$ galaxies and galaxy groups from APM and SDSS early data release.

-2005, Scranton et al. \cite{scranton05} found cross-correlations between $\sim 200,000$ QSOs and $\sim 13$ million galaxies from the SDSS ranging from {\it  positive to negative signal}, depending of the magnitude limits of the QSO population subsample ($s=\{1.95,1.41,1.07,0.76,0.50\}$).

\section{Theory: Analytical}

The most successful hypothesis to explain the observed cross-correlations and anti-correlations is gravitational lensing. 
It generates two competing effects that can explain both positive and negative cross-correlations between objects of two redshift distinct populations, in what is called magnification bias.

The presence of a gravitational lens magnifies sources behind it, bringing to view sources that would be too faint to be detected in a magnitude-limited survey. 
This effect works to produce a {\it positive} cross-correlation between objects physically associated to the foreground lenses and background objects that are magnified.
On the other hand, the lens also enlarges the solid angle behind it, therefore the source density behind the lens is diluted, what works to produce a {\it negative} background-foreground cross-correlation.

The factor that determines which of the two competing effects (magnification or dilution) is preponderant is the slope of the magnitude number count of the sources. If this slope is steep then many faint sources are brought to view, but if the slope is low (flatter) then few extra sources are brought to view and the dilution effects wins.

To compute the cross-correlation between two populations we calculate the correlation between the density contrast of the two groups
\begin{equation}
  \omega_{qg}(\theta) \equiv
  \left< \left[ \frac {n_q(\bm{\phi})}{\bar{n}_q} -1 \right]
  \left[ \frac {n_g( \bm{\phi}+ \bm{\theta})}{\bar{n}_g} -1 \right]
  \right>  \; ,
  \label{cross-correl-def}
\end{equation}
where $n_q$ and $n_g$ are the background and foreground populations (e.g. QSOs and galaxies or galaxy groups) densities. A bar over a quantity indicates its mean value, and $\left< ... \right>$ represents the average over $\bm{\phi}$ and the direction of  $\bm{\theta}$ (but not its modulus).
This definition is equivalent to the cross-correlation estimator \cite{GMS05}
\begin{equation}
  \omega_{qg}(\theta) =  \frac{DD(\theta)}{DR(\theta)} -1  \; ,
  \label{estimator}
\end{equation}
where $DD(\theta)$ is the observed number of background-foreground pairs, and 
$DR(\theta)$ is the expected number of random pairs.

The ratio $DD(\theta) / DR(\theta)$ is the enhancement factor due to the magnification bias, and under the assumption that the cumulative number counts by flux is of the form $N(>S)\propto S^{-s}$ we have that
\begin{equation}
  \omega_{qg}(\theta) = \mu(\theta)^{s-1} - 1  \; ,
  \label{w_qg-mag-rel}
\end{equation}
which cleary indicates that in an overdense region ($\mu>1$): $s>1$ leads to positive cross-correlation, and $s<1$ leads to negative cross-correlation (anticorrelation).

The magnification can be written in terms of the gravitational lensing convergence $\kappa$ and shear $\gamma$,
\begin{equation}
  \mu({\bm \theta}) = 
  \frac{1}{\left| \left[1-\kappa({\bm \theta}) \right]^2-\gamma^2({\bm \theta}) \right| } \;.
  \label{magnification}
\end{equation}
The magnification can be calculated in the cosmological context, assuming the Born approximation, as a weighted integration of the matter density field
\begin{equation}
  \kappa({\mbox{\boldmath$\theta$}}) =
  \int_0^{y_{\infty}} { W(y) \delta({\mbox{\boldmath$ \theta$}},y) dy} \; ,
  \label{convergence2}
\end{equation}
where $\delta$ is the density contrast, $y$ is a comoving distance, 
$y_{\infty}$ is the comoving distance to the horizon, and $W(y)$ is a lensing weighting function
\begin{equation}
  W(y) = \frac{3}{2} \left( \frac{H_o}{c} \right)^2 \Omega_m
  \int_y^{y_{\infty}} { \frac{G_q(y^\prime)}{a(y)}
  \frac{f_K(y^\prime-y)f_K(y)}{f_K(y^\prime)} dy^\prime } \; .
  \label{weight}
\end{equation}
$G_q$ is the source distribution, $a$ is the scale factor, and $f_K$ is the curvature-dependent radial distance.

The shear can be obtained from a convolution of the convergence \cite{bartelmann-schenider01}, and therefore the knowledge of the mass distribution between the observer and the source plane allows the computation of the desired gravitational lensing effects.

If we assume weak gravitational lensing, $\kappa \ll 1$, some analytical calculations become much simpler. 
The magnification (\ref{magnification}) becomes $\mu = 1+2\kappa$ and the QSO-galaxy cross-correlation (\ref{cross-correl-def}) can be expressed as \cite{dolag_bartelmann97,GBB01}
\begin{eqnarray}
  \omega_{qg}(\theta) & = &
    {\displaystyle \frac{(s-1)}{\pi} \frac{3}{2}
    \left( \frac{H_o}{c} \right)^2 \Omega_m }
    {\displaystyle \int _{0}^{y _{\infty} }} dy
    {\displaystyle \frac { W_g(y) \, G_q(y) } {a(y)}}  \nonumber \\
    & &  \times
    {\displaystyle \int _{0}^{\infty }} dk \, k \,
    P_{gm}(k,y )\, J_0[f_K(y )k\theta ]  \, ,
         \label{analytic-cross-correl}
\end{eqnarray}
where $y$ is the comoving distance, which here parameterizes time
($y_{\infty}$ represents a redshift $z=\infty$),
and $k$ is the wavenumber of the density contrast in a plane wave
expansion;
$J_0$ is the zeroth-order Bessel function of first kind;
and $f_K(y)$ is the curvature-dependent radial distance ($=y$ for a flat
universe).
$P_{gm}(k,y )$ can be seen as the galaxy-mass cross-power spectrum \cite{jain03}, and under some assumptions \cite{GBB01} may be expressed as 
$P_{gm}(k,y )=\sqrt{P_g(k) P_m(k,y )}$, where
$P_g(k)$ is the power spectrum for galaxies or galaxy groups and
$P_m(k,y)$ is the non-linear time evolved mass power spectrum.

\begin{figure}
  \centering
  \includegraphics[width=8.7 cm]{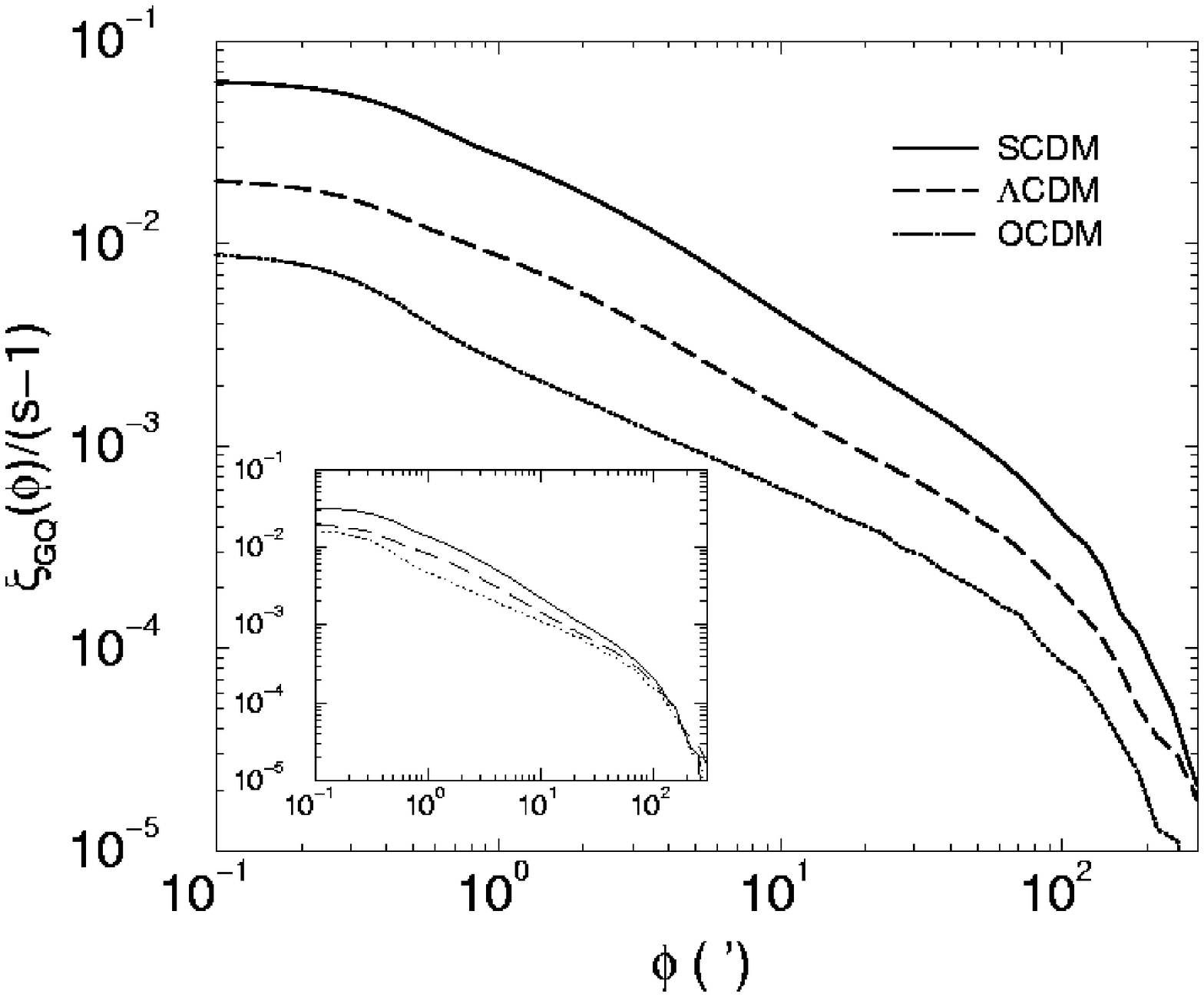}
  \includegraphics[width=8.7 cm]{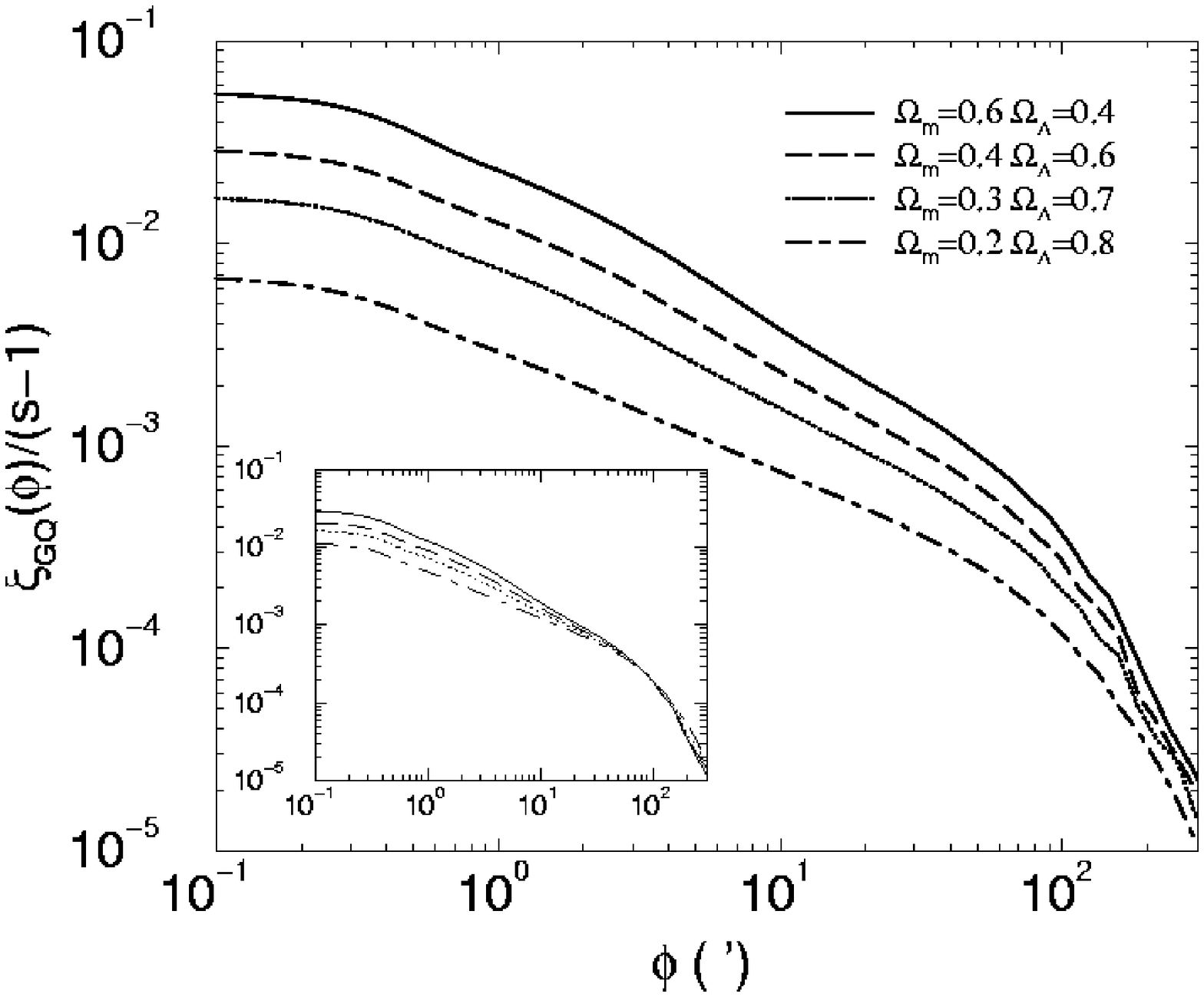}
  \vspace{0.cm}
  \caption{\label{cosmo-depend}   
    Cross-correlation due to weak gravitational lensing dependence on cosmological model (top plot) and matter density (bottom plot). The internal plots are the results for the mass power spectrum normalized to the cluster abundance (main curves use COBE normalization). 
{\it Top plot}: Solid lines are for $SCDM$, $\Omega_m=1$, $h=0.5$ ($\sigma_8=1.1$); 
dashed ones for $\Lambda CDM$, $\Omega_m=0.3$, $\Omega_{\Lambda}=0.7$,
$h=0.7$ ($\sigma_8=1.0$); and
dotted ones for $OCDM$, $\Omega_m=0.3$, $\Omega_{\Lambda}=0$, $h=0.7$ 
($\sigma_8=0.46$). 
{\it Bottom plot}: Dependence on matter density in a flat universe with
  cosmological constant ($\Omega_m + \Omega_{\Lambda}=1$). 
Solid lines are for $\Omega_m=0.6$ ($\sigma_8=1.4$); 
dashed ones for $\Omega_m=0.4$ ($\sigma_8=1.2$);
dotted ones for $\Omega_m=0.3$ ($\sigma_8=1.0$); and
dot-dashed ones for $\Omega_m=0.2$ ($\sigma_8=0.72$).
Other parameters are 
$\Omega_b=0.019/h^2$, $h=0.7$, $n=1$
  }
\end{figure}

Expression (\ref{analytic-cross-correl}) indicates that the background-foreground cross-correlation due to lensing is dependent on several quantities of cosmological relevance. 
Guimar\~aes et al. \cite{GBB01} explores these cosmological dependences and Figure \ref{cosmo-depend} illustrates the sensitivity of the cross-correlation between a population of QSOs at z=1 and galaxies at z=0.2.

Higher order terms can be added to the Taylor expansion of the magnification \ref{magnification}, producing a better approximation \cite{menard03}.
However the full accounting of non-linear magnification is not feasible by this analytical path.

\section{Simulations}

The analytical approach has some limitations at fully incorporating deviations from the weak lensing approximation and properly modeling lensing selection issues.
Therefore, computer simulations can be a useful complement for a better understanding of the problem.
We describe, in the sequence, two kinds of simulations: one aiming at very large regions and many lenses, collectively analyzed, but somewhat short on resolution on small scales, and another kind of simulation aiming at individual clusters and their substructure.

\subsection{Galaxies and Galaxy Groups}
Both the mass and light (galaxy) distributions in the universe can be mocked from N-body simulations, and from those the cross-correlation between background and foreground objects due to gravitational lensing can be obtained \cite{GMS05}.

The first step is to generate a representation of the mass distribution in a redshift cone from the observer ($z=0$) to a source plane at high redshift. 
A galaxy mock catalog can be generated from the simulated density field and the adoption of a bias prescription for the galaxy population. This galaxy mock will have the galaxy density and auto-correlation function desired, which can be set to mimic a chosen real galaxy survey.
Also from the simulated density field the gravitational lensing for a chosen source plan can be calculated in the form of a lensing map (convergence, shear, or magnification) using the formalism of the previous Section.

Guimar\~aes et al. \cite{GMS05} used the Hubble Volume Simulation, a N-body simulation with $10^9$ particles of mass
$M_{part}=2.25 \cdot 10^{12}h^{-1}\msun$ in a periodic
$3000^3 h^{-3}\rm{Mpc}^3$ box, initial fluctuations generated by CMBFAST, force resolution of $0.1h^{-1}\rm{Mpc}$, and a ``concordance model'' parameter set, $\Omega_M=0.3$, $\Omega_\Lambda=0.7$, $\Gamma=0.21$, $\sigma_8=0.90$.
The magnification map was calculated for a source plane at redshift 1, and the average magnification was measured around chosen sets of lenses (galaxies or galaxy groups of varying membership identified in the projected sky mock) to determine the expected source-lens cross-correlation.

\begin{figure}
  \centering
  \includegraphics[width=8.2 cm]{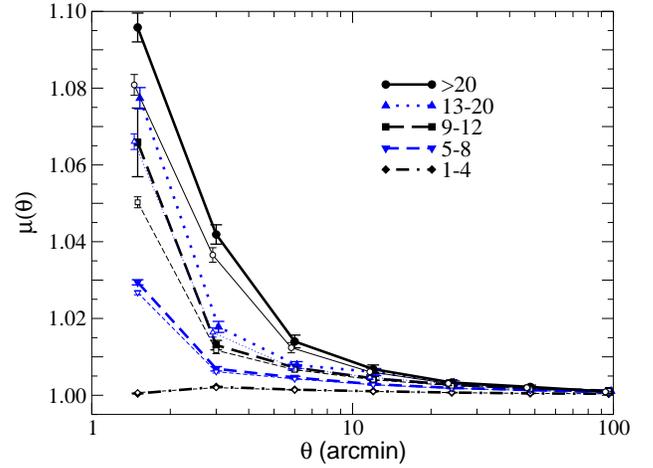}
  \caption{\label{mag_groups}  
    Average magnification around mock galaxy groups. The number ranges in the legend are the number of galaxies for the sets of groups. Thick lines with filled symbols include departures from the weak lensing approximation, and thin lines with open symbols are the weak lensing approximation for the magnification calculation. Errors are the standard deviation of the mean.}
\end{figure}

\begin{figure}
  \centering
  \includegraphics[width=8.2 cm]{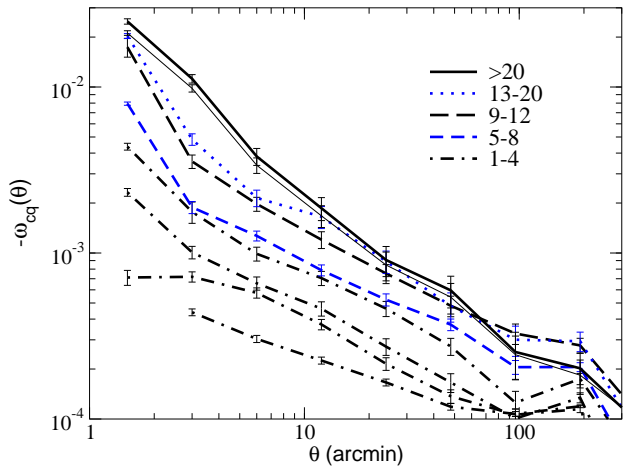}
  \caption{\label{cross-correl_groups}
    Cross-correlation between QSO and mock galaxy groups. 
The number ranges in the legend are the number of galaxies for the sets of groups. Thick lines include departures from the weak lensing approximation, and thin lines are the weak lensing approximation for the magnification calculation. Curves for groups of membership 1 to 4 are shown individually (from bottom to top for low to high membership). Errors are the standard deviation of the mean.}
\end{figure}

Figure \ref{mag_groups} shows the average magnification around galaxy groups, and Figure \ref{cross-correl_groups} shows the corresponding cross-correlation. Groups of larger membership trace denser regions, so have a higher magnification and stronger cross-correlation amplitude.

\begin{figure}
  \centering
  \includegraphics[width=8. cm]{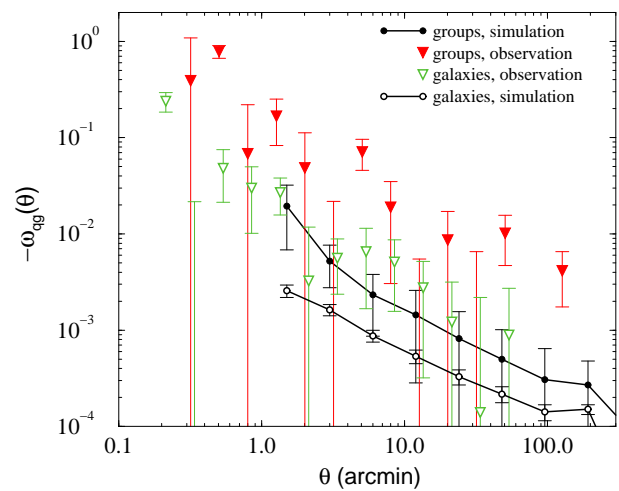}
  \caption{\label{cross-correl_obs}
    QSO-galaxy and QSO-group cross-correlation.
Observational data is from Myers et al. \cite{myers03,myers05} with field-to-field errors.
Simulation uses groups with 9 or more galaxies and estimated error with same source density as observed data. Some observational points fall below the shown logarithmic scale. }
\end{figure}

Figure \ref{cross-correl_obs} compares the simulation results for galaxies and galaxy groups with observational data from Myers et al. \cite{myers03,myers05}. 
Simulation results for angles smaller than 1 arcmin cannot be obtained due to limited simulation resolution; however for angles from 1 to 100 arcmin the comparison with data gives a large disagreement between the amplitudes of observed and simulated cross-correlations.

This disagreement between some observed QSO-galaxy and QSO-group cross-correlations were already visible in Figure \ref{observations} and is source of controversy.

\subsection{High Resolution Galaxy Clusters}

One limitation of the simulations described in the previous Section is the low resolution at small scales.
To be able to probe small angular scales, and therefore regions near the cluster core or the halo substructure, it is necessary to use simulations of much higher resolution.

Guimar\~aes et al. \footnote{work in preparation, to be published in detail elsewhere} used high resolution cluster halo simulations carried out by the Virgo Consortium \cite{navarro04} to study the gravitational lensing magnification due to galaxy cluster halos. 
These halos were generated by high mass resolution resimulations of massive halos selected from a large $(479\mpc)^3$ N-body simulation.

The magnification maps generated by the simulated cluster halos were calculated using equation (\ref{magnification}), assuming a source plan at redshift 1 and the cluster halo at redshift 0.15.
To evaluate the role of substructure it was also calculated the magnification of the homogenized halo in concentric rings, so the density profile is maintained, but the substructure is washed away.
The weak gravitational lensing approximation was also used to calculate the magnification, so the departure from this regime can be quantified in the case of massive clusters.

\begin{figure*}
  \includegraphics[width=15 cm]{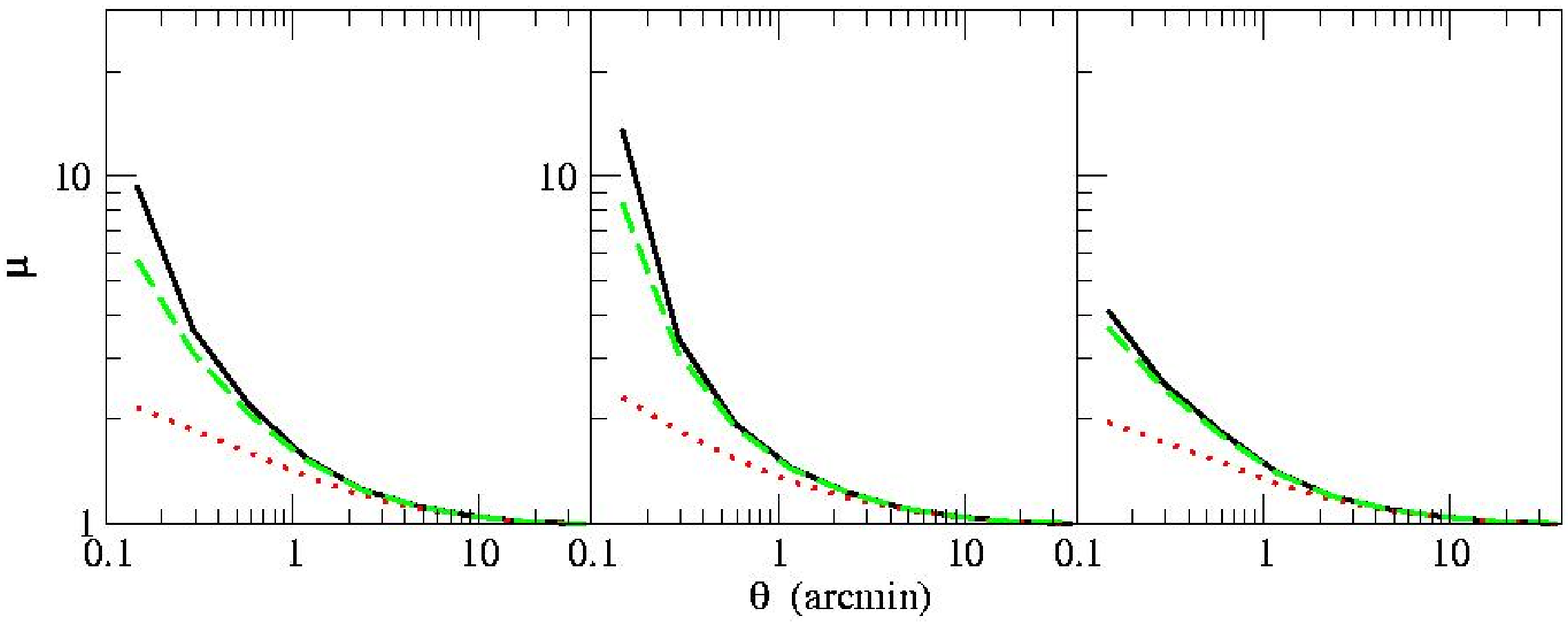}
  \includegraphics[width=15 cm]{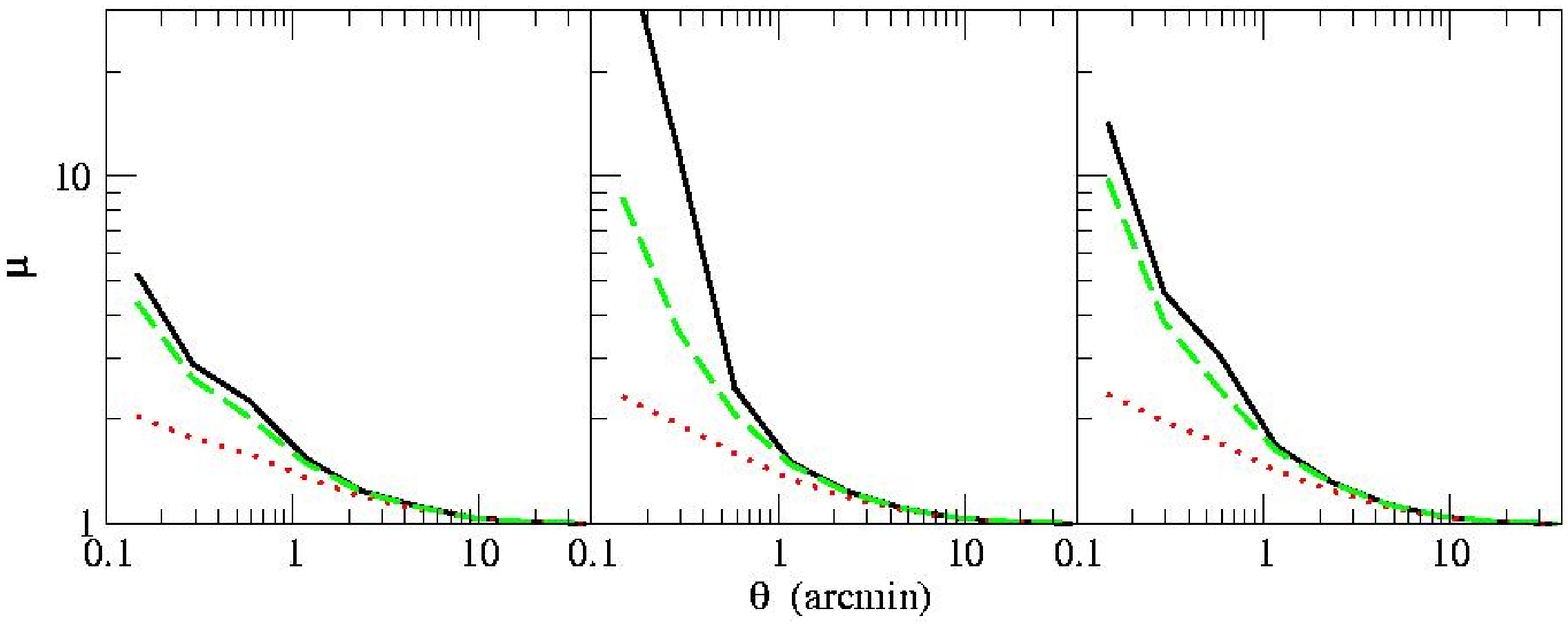}
  \includegraphics[width=15 cm]{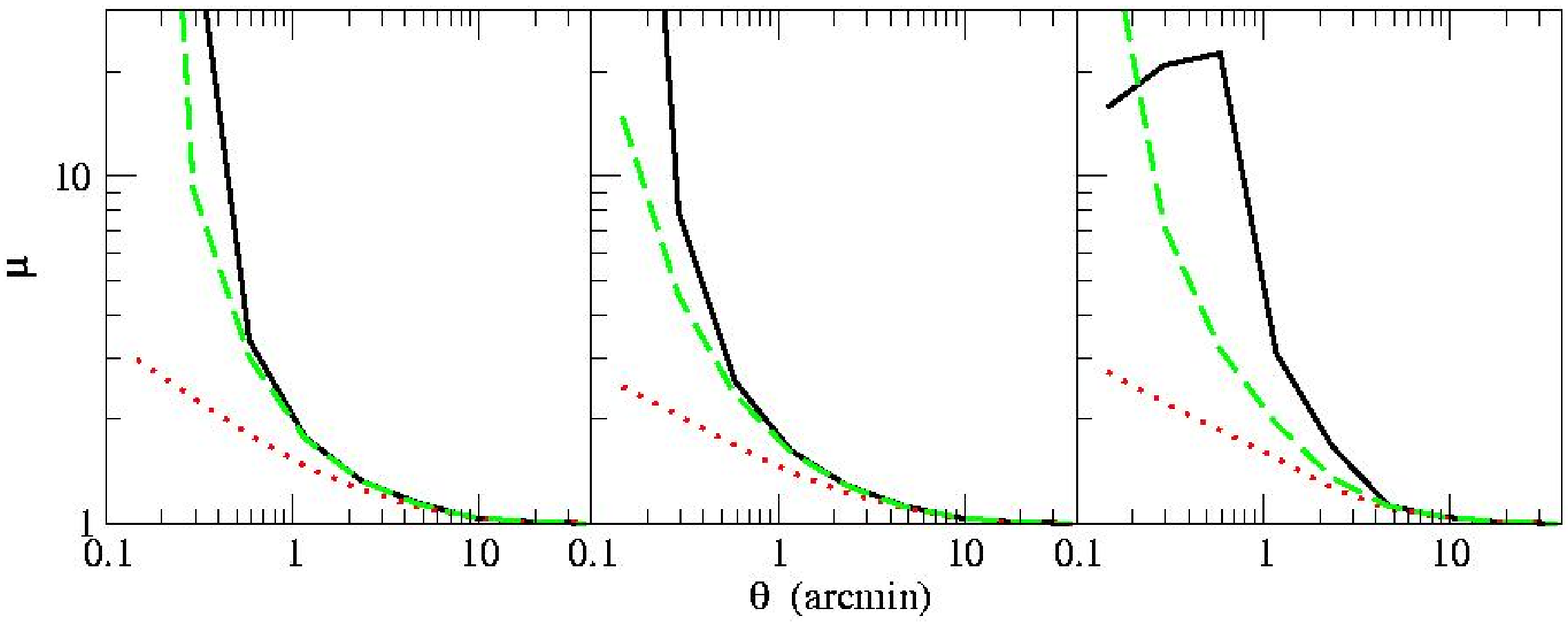}
  \caption{\label{clusters} 
    Magnification curves for three simulated clusters viewed from three orthogonal directions. {\it Solid curves} take in account the non-linear magnification due to cluster substructure; {\it dashed curves} are for the homogenized mass  inside concentric rings; {\it dotted curves} use weak lensing approximation. }
\end{figure*}

Figure \ref{clusters} shows for three cluster halos seen by three orthogonal directions each, the average magnifications described above as a function of the angular distance to the halo center.
Five other simulated clusters examined show similar curves (results not presented).

Departures from the weak lensing regime become important at angles of few arcmin. 
At these same scales the contribution of substructure to the magnification can also be significant in some cases.

\section{Discussion}

We reviewed some of the accumulated history of QSO-galaxy cross-correlation observations, and the gravitational lensing theory associated with it.

The observations of cross-correlation in the sky between populations of objects that are very apart in depth are old, numerous, and are becoming precise.
The gravitational lensing explanation for the cross-correlation phenomenon is more recent and carries with it the possibility of using these kind of measurements as a tool for cosmology and astrophysics.
However, even though the magnification bias hypothesis is in qualitative agreement with observations in general, several measurements of the QSO-galaxy cross-correlation have a higher amplitude than what is predicted.

The most recent and largest measurements of QSO-galaxy cross-correlation carried out using the approximately 200,000 quasar and 13 million galaxies of the Sloan Digital Sky Survey (SDSS) \cite{scranton05} are in agreement with theoretical predictions based on a ``concordance model'' and weak gravitational lensing (non-linear magnification was not taken into account by \cite{scranton05}, but is considered to be necessary by the authors for a more accurate modeling).

On the observational side, further work with the existing data may clarify the reason for disagreements among different measurements of QSO-galaxy cross-correlation. 
If systematic errors are to blame for the huge amplitudes measured by various groups, then it is fundamental to identify and characterize them.

On the theoretical side, both analytical work and the use of simulations are helping to provide a realistic description of the phenomenon. 
This theoretical framework in conjunction with observational data may be useful in determining quantities of astrophysical and cosmological interest, for example and most promisingly the average mass of lens populations and the galaxy-mass power spectrum.

\bibliography{100year-1}

\end{document}